 \documentclass[11pt]{article}

 \pdfoutput=1   % for activate  PDFLATEX

\usepackage{graphicx}
\usepackage{amsmath}

\usepackage{enumitem}

\usepackage[francais,english]{babel}

\usepackage{float}
\usepackage{ifthen}
\usepackage{color}

\usepackage{hyperref}

\begin{document}

%-------------------------------------------------------------------

\title{On the concept of pseudo-energy of T. G. Shepherd.\\
{\it A close link with \underline{exergy} functions\/}.}

\author{by Pascal Marquet. {\it M\'et\'eo-France.}}

%\corraddr{Pascal MARQUET, DPr\'evi/Labo, M\'et\'eo-France, 42 av. G. Coriolis, 31057 Toulouse CEDEX 01, France.\\ Web site: http://perso.numericable.fr/$\sim$pmarquet/  ; E-mail: pascal.marquet@meteo.fr }

\date{\today}
%\date{10th of January, 2014}
%\vspace*{-2mm}
\maketitle

%-------------------------------------------------------------------

\vspace*{-10mm}

\begin{center}
{\em Annotated, corrected and augmented copy of a Note submitted in January 1994 to the \underline{Quarterly Journal of the Royal Meteorological Society} (revised in June 1994).} \\
{\em Published in Volume 121, Issue 522, pages 455--459, January 1995 Part B}: \\
\url{http://onlinelibrary.wiley.com/doi/10.1002/qj.49712152212/abstract} \\
{\em \underline{Corresponding address}: pascal.marquet@meteo.fr}
\end{center}
\vspace{1mm}

% -----------------------
 \section{INTRODUCTION.} % (Section 1)
% -----------------------
\label{section_1}

In a recent paper, Shepherd (1993) derived a general expression for the available potential energy for compressible, hydrostatic flow, where the sum of this available energy and the kinetic energy is called pseudo-energy. 
He demonstrated that for the special choice of a basic state defined by $\theta_0(p) = \overline{\theta}(p)$ where the potential temperature $\overline{\theta}(p)$ is the average on an isobaric surface, the small-amplitude limit of the generalized available potential energy reduces to the well-known approximate form of Lorenz (1955) expressed in a pressure vertical coordinate.

But other forms of available energies exist in atmospheric energetics and in thermodynamics, where the name {\it exergy\/} has been coined by Rant (1956) to denote the maximum work that can be extracted from any system when it is subject to some constraints (adiabatic transformations or constant total energy for instance).

The purpose of this note is to show that the specific available enthalpy function which is the flow energy of a fluid -- denoted by $a_h = (h - h_r) - T_r \: (s - s_r)$ see Marquet (1991) -- can be obtained from the generalized approach of Shepherd if a constant basic state at temperature $T_r$ and pressure $p_r$ is considered. 
This special form of pseudo-energy also leads to the global hydrostatic concepts of Dutton (1973) and Pichler (1977). The function $a_h$ only depends on the specific enthalpy $h$ and entropy $s$ at any point, the values $h_r$ and $s_r$ refer to the special dead state at temperature $T_r$ and pressure $p_r$.

It is also explained that for a real isothermal basic state made up of an atmosphere at constant temperature $T_0$ but with a variable pressure, the generalized expression of Shepherd reduces with a good accuracy to the approximate functions introduced by Pearce (1978) or Blackburn (1983) in meteorology, it is moreover exactly the primary result obtained by Thomson (1853) in thermodynamics.

% -----------------------
 \section{SHEPHERD'S (1993) PSEUDO-ENERGY.} % (Section 2)
% -----------------------
\label{section_2}

The theory of Shepherd will not be described in detail, only the main steps will be sketched.
We start with the non-canonical Hamiltonian representation of the system written in the so-called symplectic form
\begin{equation}
\frac{\partial v_i}{\partial t} 
\; = \; 
\mbox{\boldmath $I$}_{ij} \: 
\frac{\delta {\cal H}}{\delta v_j} 
\: ,
\label{eq_1}
\end{equation}
where the evolution of the dynamical variables ($v_i$) depends on the skew-symmetric Poisson tensor ($\mbox{\boldmath $I$}_{ij}$) and on the functional derivative of the Hamiltonian  (${\cal H}$)\footnote{\color{blue} 
Additional explanations may be found in McIntyre and Shepherd (1987) and Shepherd (1990a, 1990b, 2003).}.

Euler's equations for a compressible, hydrostatic, perfect gas correspond to the following non-canonical Hamiltonian
\begin{equation}
\left.
\begin{aligned}
{\cal H} & \; = \;  E_k \: + \: H 
\; \equiv \; \iiint_{\cal M} \left[ \; (\vec{u})^2/2 \: + \: c_p \: T \; \right] \; dm \\
{\cal H} & \;
\; \equiv \; \iiint_{\cal M} \left[ \; (\vec{u})^2/2 \: + \: c_p \; \theta \; \Pi(p) \; \right] \; dm
\end{aligned}
\;\; \right\}
\label{eq_2}
\end{equation}
where $m$ is an element of mass of the atmosphere and ${\cal M}$ is the mass integrating domain of the atmosphere.\footnote{\color{blue} This formulation for the Hamiltonian is given in Eq.(8.1) of Shepherd (1993).} 
The specific kinetic energy of the horizontal wind $(\vec{u})^2/2$ is used because of the hydrostatic hypothesis (thus by dropping the vertical wind component).
The integral of $(\vec{u})^2/2$ is the total kinetic energy which is denoted by $E_k$.
The specific enthalpy for the dry perfect gas is supposed to be $c_p \:T = c_p \; \theta \; \Pi(p)$, where $c_p = 1004$~J~K${}^{-1}$~kg${}^{-1}$ is the specific heat of dry air at constant pressure, $\Pi(p)= T \: / \: \theta = (p/p_{00})^{\kappa}$ is the Exner's function with $p_{00}= 1000$~hPa and $\kappa = R/c_p$, where $R = 287$~J~K${}^{-1}$~kg${}^{-1}$ is the gas constant of dry air. 
The global enthalpy is denoted by $H$.

The pseudo-energy is defined by Shepherd as\footnote{\color{blue} 
The first formulation ${\cal H}( \vec{v} ) - {\cal H}( \vec{V} ) + {\cal K}( \vec{v} ) - {\cal K}( \vec{V} )$ is given in Eq.(5.8) of Shepherd (1993).
The second formulation $( {\cal H} + {\cal K}) (\vec{v}) \: - \:  ( {\cal H} + {\cal K}) (\vec{V})$ is given in Eq.[35] of Shepherd (2003).}
\begin{equation}
\boxed{ \; \; 
{\cal A}
\; = \; 
{\cal H}( \vec{v} )
\: - \: 
{\cal H}( \vec{V} )
\: + \: 
{\cal K}( \vec{v} )
\: - \: 
{\cal K}( \vec{V} ) 
\; \; }
\;
\; = \; 
( {\cal H} + {\cal K}) (\vec{v})
\: - \: 
( {\cal H} + {\cal K}) (\vec{V})
\: 
\label{eq_3}
\end{equation}
where $\vec{V}$ is a resting basic state in terms of the state vector $\vec{v}$ of the Hamiltonian ${\cal H}$ of system (\ref{eq_1}).
Even if the Hamiltonian ${\cal H}(\vec{V})$ corresponds to an equilibrium state, ${\cal H}(\vec{v}) - {\cal H}(\vec{V})$ is only {\it linear\/} with respect to the perturbation amplitude $\delta \vec{v} = \vec{v} - \vec{V}$, and it was necessary to introduce some
{\it Casimir invariant\/} denoted by ${\cal K}$ in order to make ${\cal A}$ quadratic with respect to this disturbance $\delta \vec{v}$.
%===============
% Figure (1):
%===============
\begin{figure}[t]
\centering
\includegraphics[width=0.99\linewidth,angle=0,clip=true]{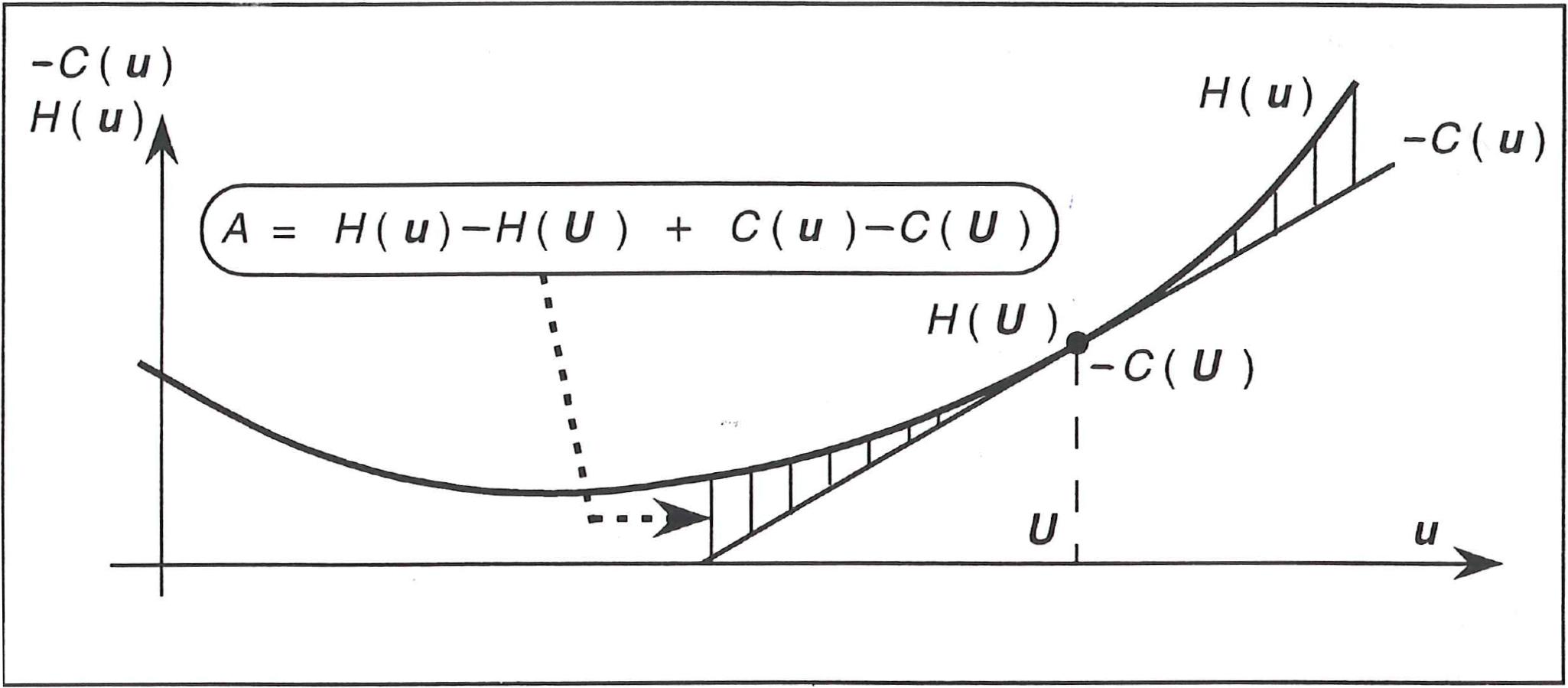}
\caption{
{\it
A one dimensional vision of Hamiltonian ($H$), Casimir invariants ($C$) and pseudo-energy ($A$) depending on a state variable ``$u$''.
This Figure is from Marquet (1994).
Since $H(u)$ is increasing for $u$ close to $U$, the Hamiltonian is not quadratic with respect to ($u-U$).
If ``$-\:C(u)$'' defines the tangent of the Hamiltonian $H(u)$ at point $u=U$, then the pseudo-energy defined by $A(u) = [ \: H(u) - H(U) \:] + [ \: C(u) - C(U) \:]$ is represented by the dashed area and is clearly quadratic with respect to ($u-U$).
}
\label{Fig_1}}
\end{figure}

The Casimir invariants somehow span the tangent manifold of ${\cal H}$  at point $\vec{V}$ (see Fig.\ref{Fig_1}).\footnote{\color{blue} 
This Figure was in Marquet (1994) and was not published in Marquet (1995).}
Since the Casimir are ``invariant'', they are solution of ${\partial v_i}/{\partial t} = 0$ and from (\ref{eq_1}) the Casimir ${\cal K}$ are thus defined by
\begin{equation}
\left.
\mbox{\boldmath $I$}_{ij} \: 
\frac{\delta {\cal K}}{\delta u_j} 
\right|_{\vec{V}}
\; = \; 
0 
\;  \;
\Rightarrow
\; \; 
\left.
\frac{\delta {\cal H}}{\delta u_j} 
\right|_{\vec{V}}
\: = \; 
 - \: 
\left.
\frac{\delta {\cal K}}{\delta u_j} 
\right|_{\vec{V}}
\: .
\label{eq_4}
\end{equation}
Solutions of $\delta {\cal K} / {\delta u_j}|_{\vec{V}} \neq 0$ exist because the Hamiltonian is non-canonical, i.e. $\mbox{\boldmath $I$}_{ij} $ is non-invertible.

Following Shepherd (1993) the Casimir invariant of Eq.(\ref{eq_2}) is the integral over the mass ${\cal M}$ of a function $C(\theta)$ which only depends on the potential temperature $\theta$. 
The function $C(\theta)$  is determined from Eq.(\ref{eq_2}) by the functional derivative of the Hamiltonian $\delta {\cal K} / {\delta \theta} = c_p \: \Pi(p)$ together with the
second part of Eq.(\ref{eq_4}) applied to the resting basic state $\vec{V}$ which will be denoted by a subscript ``$0$'' hereafter.
It can be inferred that $dC(\theta)/d\theta = - \: \delta {\cal H} / {\delta \theta} = - \: c_p \: \Pi(p)$ and that the Casimir are equal to\footnote{\color{blue} These formulations $dC(\theta)/d\theta= - \: c_p \: \Pi(p)$ and (\ref{eq_5}) for ${\cal K}$ are given in Equations (8.8) and (8.9) of Shepherd (1993).} 
\begin{equation}
{\cal K} \; = \; 
    - \:  \iiint_{\cal M} \left[ 
       \; \int^{\theta} \; c_p 
       \; \Pi\left\{ {\cal P}(\theta')\right\} 
       \; d\theta' \; 
          \right] \; dm
  \: .
\label{eq_5}
\end{equation}

It is assumed that $\partial \theta_0 / \partial p < 0$ so that the inverse function ${\cal P}[\:\theta_0(p)\:] = \theta_0^{-1}(p)$ exists. 
After some manipulation it results from Eq.(\ref{eq_3}) with the use of Eqs.(\ref{eq_2}) and (\ref{eq_5}) that 
\begin{equation}
{\cal A}
\; = \; 
\left( {\cal H} \: - \:  {\cal H}_0 \right)
\: + \: 
\left( {\cal K} \: - \:  {\cal K}_0 \right)
\; = \; 
\iiint_{\cal M} 
\left[ 
    \; (\vec{u})^2/2 \: + \: a \; 
\right] 
\; dm  \: ,
\label{eq_6}
\end{equation}
or equivalently that
\begin{equation}
{\cal A}
\; = \; 
E_k
\; - \; c_p \;
\iiint_{\cal M} 
\left[ 
   \; \int_0^{\theta-\theta_0}
   \left[ \;
   \; \Pi\left\{ {\cal P}(\theta''+\theta_0)\right\} 
   \; - \: 
   \; \Pi\left\{ {\cal P}(\theta_0)\right\} 
   \; \right] 
   \; d\theta'' \; 
\right] 
\; dm
\: ,
\label{eq_7}
\end{equation}
where the variable $\theta'' = \theta'' - \theta_0$ is used.\footnote{\color{blue} 
This formulation is given in Eq.(8.10) of Shepherd (1993).}

Equation (\ref{eq_7}) gives the pseudo-energy as expressed by Shepherd. 
According to Eq.(\ref{eq_6}) the integrand $(\vec{u})^2/2 \: + \: a \:$ is the specific form of the pseudo-energy. 
It is the sum of the specific kinetic energy $e_k = (\vec{u})^2/2$ and the ``specific available potential energy'' which can be written as
\begin{equation}
a
\; = \; - \; c_p \;
   \; \int_0^{\theta-\theta_0}
   \left[ \;
   \; \Pi\left\{ {\cal P}(\theta''+\theta_0)\right\} 
   \; - \: 
   \; \Pi\left\{ {\cal P}(\theta_0)\right\} 
   \; \right] 
   \; d\theta''
\: .
\nonumber
\end{equation}
Shepherd mentions that both $e_k$ and ``$a$'' obey local conservation laws, i.e. they can be interpreted locally as real forms of energy attached to a parcel of the fluid.\footnote{\color{blue} 
More precisely, it is mentioned in Shepherd (1993) that the integrand ($A = e_k + a $) of (\ref{eq_6}) ``{\it represent the pseudoenergy per unit mass, and is guaranteed to obey a local conservation law --  given by Eq.(8.11) -- of the form\/} $\partial A / \partial t + \overrightarrow{\nabla} . \, \vec{F} = 0$ {\it for some flux \/}$\vec{F}\,$''.}

% -----------------------
 \section{EXERGY FUNCTIONS.} % (Section 3)
% -----------------------
\label{section_3}

According to Shepherd's paper (Eq.(8.12) in section 8), Lorenz's available potential energy corresponds to the leading order (small-amplitude limit) approximation of Eq.(\ref{eq_7}).
Indeed the term 
$\Pi\left\{ {\cal P}(\theta''+\theta_0)\right\}  - \Pi\left\{ {\cal P}(\theta_0)\right\}$
can be approximated by the quadratic function\footnote{\color{blue} 
See the integrand of Eq.(8.12) in Shepherd (1993)}
\begin{equation}
a
\; \;  \approx \; \;
\frac{R}{p} \: \left( \frac{p}{p_{00}} \right)^{\kappa} \:
\frac{1}{(- \: d\theta_0/dp)} \; \;  \frac{(\theta - \theta_0)^2}{2}
\: .
\nonumber
\end{equation}
Clearly, the special choice $\theta_0(p) = \overline{\theta}(p)$ gives the integrand of the global available potential energy (APE) of Lorenz (1955).
This close link between the APE of Lorenz and the pseudo-energy ($a$) of Shepherd demonstrates the importance of both of them.
This can be understood as a kind of cross-validation of the two concepts. 
However, the same limitation $d\theta_0/dp \neq 0$ exists for the two concepts if $\theta_0 (p) = \overline{\theta}(p)$ which needs the definition of the inverse function ${\cal P}[\:\theta_0(p)\:] = \theta_0^{-1}(p)$.
This appears to be a drawback and a lack of generality.

The APE of Lorenz is not the only way to define available energy in atmospheric science.
It is possible to use {\it exergy\/} functions defined by the founders of thermodynamics (Thomson 1853, Gibbs 1873) to define the static entropic energy (Dutton 1973, Pichler 1977), the available energy (Pearce 1978, Blackburn 1983), the potential energy\footnote{\color{blue} The potential energy of Andrews (1981) was not mentioned in Marquet (1995).} (Andrews, 1981) or the moist available enthalpy (Marquet, 1991, 1993, 1994, 2003a, 2003b).

If these {\it exergy\/} functions are to be relevant to atmospheric energetics, and if the pseudo-energy is a general concept, {\it exergy\/} functions must be similarly derived as special cases of the pseudo-energy of Shepherd.
These results have not been derived by Shepherd and they will be demonstrated in this section by choosing another definition for $\theta_0(p)$.

To demonstrate existing links between pseudo-energy and available enthalpy, it is easier to start with an equation different from (\ref{eq_7}) in order to avoid the hypothesis $d \theta_0 /dp \neq 0$  and the use of any inverse function ${\cal P}$ of $\theta_0 (p)$.
The following expression of the pseudo-energy is directly expressed in terms of the temperatures $T = \theta \: \Pi(p)$ and $T_0 = \theta_0 \: \Pi_0(p)$, and in terms of the potential temperatures $\theta$ and $\theta_0$:
\begin{equation}
{\cal A}
\; = \; 
E_k
\; + \;
\iiint_{\cal M} 
\left[ \;
     c_p \; ( T - T_0 )
   \: - \: c_p
        \int^{\theta}_{\theta_0}
             \frac{T_0(\theta')}{\theta'}
      \; d\theta' \; 
\right] 
\; dm
\: .
\label{eq_8}
\end{equation}
It is possible to derive this equation as an intermediate step in computations of Shepherd's paper.
It is obtained by avoiding the use of the inverse function ${\cal P}$.
The pseudo-energy ${\cal A}$ as defined by (\ref{eq_8}) can be computed as far as the resting basic state defined by $T_0(\lambda, \varphi, \theta)$, and $\theta_0(\lambda, \varphi, p)$ is known at every point located at $(\lambda, \varphi, p)$ where $\lambda$  and $\varphi$ are latitude and longitude respectively.

Starting from Eq.(\ref{eq_8}) the special case $T_0 = T_r = c_1^{\rm{ste}}$ and $\theta_0 = \theta_r = T_r \: (p_r/p_{00})^{\kappa} = c_2^{\rm{ste}}$ can be studied, whereas it is not possible starting from Eq.(\ref{eq_7}) since the reference potential temperature is constant and the inverse function ${\cal P} = \theta_r^{-1}$ does not exist.
The values of the two constants,  $c_1^{\rm{ste}}$ and $c_2^{\rm{ste}}$, will not be determined. 
This basic state is not a reference ``atmosphere'', it is made up of a fixed parcel of fluid at constant temperature $T_r$ and pressure $p_r$.
 It is the one used in some study of available energy (Pearce, 1978) and in the study of specific available enthalpy (Marquet, 1991).
The same definition was already suggested by Gibbs (1873, page~54) for defining the concept of available energy in the global atmosphere.\footnote{\color{blue} 
The old concept of availability in energy defined by Thomson (Lord Kelvin), Gibbs and others are described with some details in Marquet (1991), see \url{http://arxiv.org/abs/1402.4610} {\tt arXiv:1402.4610 [ao-ph]}.}

The pseudo-energy can then be computed with Eq.(\ref{eq_8}), and as expected the result can be written as ${\cal A} = E_k + A_h$, where $A_h = A_T + A_p$ is the global available enthalpy:
\begin{equation}
\boxed{\; \;
A_h
\; = \; 
\iiint_{\cal M} 
     \;  \left[ \;   
      ( h - T_r \: s ) - (h_r - T_r \: s_r)
    \; \right] 
\; dm
\; = \; 
\iiint_{\cal M} 
     \;  \left[ \;   
      ( h - h_r ) - T_r \: (s - s_r)
    \; \right] 
\; dm
\;\;}
\: .
\nonumber
\end{equation}
The first integral can be rewritten as the integral of $\:(h+C)-(h_r+C_r)$ where $\boxed{\:C(\theta) = -\: T_r \: s(\theta) }\,$.
According to Eq.(\ref{eq_6}), the enthalpy $h$  and $C(\theta)$ can thus be interpreted in terms of an Hamiltonian plus a Casimir invariant.
Since $T_r$ is a constant reference temperature, $C(\theta)$  varies like the local entropy $s(\theta)$ which depends on the potential temperature $\theta$.
Therefore, the conservative feature observed by the Casimir $C(\theta) = -\: T_r \: s(\theta)$ is due to adiabatic and isentropic processes.

The second integral can be computed by using $h=h_r + c_p \: (T - T_r)$ and $s=s_r + c_p \: \ln(T /T_r)$, leading to
\begin{equation}
\boxed{\; \;
A_h
\; = \; 
\iiint_{\cal M} 
     c_p \; 
    \left[ \;   
      ( T - T_r ) - T_r \: \ln(T/T_r)
    \; \right] 
\; dm
\; + \:
\iiint_{\cal M} 
     R \; T_r \: \ln(p/p_r)
\; dm
\;\;}
\: .
\label{eq_9}
\end{equation}
The first integral of (\ref{eq_9}) is the temperature component $A_T$ and it is precisely the formulation obtained by Thomson (1853), latter Lord Kelvin.\footnote{\color{blue} 
See Marquet (1991) \url{http://arxiv.org/abs/1402.4610} {\tt arXiv:1402.4610 [ao-ph]}.}
The second integral is the pressure component $A_p$ which has been studied my Margules (1901) in a paper published before the more famous one of 1905 ({\it On the energy of storms\/}) which inspired Lorenz (1955).

Moreover it can be shown that Eq.(\ref{eq_9})  corresponds to the global hydrostatic version of the static entropic energy of Dutton (1973), denoted by $T_0 \: \Sigma$, or to the integral of the local associated version of Pichler (1977), denoted by $T_0 \: \sigma$. 
In these studies dealing with ``static entropic energy'' the basic state is an isothermal stratified atmosphere at $T_0$ with a density $\rho_0(z) = \rho_0(0) \: \exp[\: -g\:z/(R\:T_0) \:]$, where $g$ is gravity and $z$ is height.

The function $T_0 \: \Sigma$ is the integral over the mass ${\cal M}$  of  
$c_v \: [ \: (T - T_0)  \: - T_0 \: \ln(T/T_0)  ] - R \: T_0 \: \: \ln[\: \rho_0 (z)/ \rho \: ]$, where $c_v$ is the specific heat at constant volume.
This integral can be transformed, with the previous exponential form of $\rho_0(z)$ and with the specific internal energy $e_p = \;$geopotential$\; = \phi = g\: z$,  into the integral of $c_p \: [ \: (T - T_0)  \: - T_0 \: \ln(T/T_0)  ] + R \: T_0 \: \: \ln[\: p/ p_r \: ]$. 
This integral is thus the available enthalpy $A_h$ given by (\ref{eq_9}), provided that $p_r = p_{00} / e \approx 368$~hPa.

The hydrostatic equality between the global potential energy, internal energy and enthalpy, that is to say $E_p + E_i = H$, together with $p_0(0) \approx  p_{00} \approx e \; p_r$, have been used to obtain $T_0\:\Sigma = A_h$. 
The hydrostatic static entropic energy of Dutton is therefore closely related to the global available enthalpy $A_h$, thus to the pseudo-energy of Shepherd.

It is also interesting to examine the case of a basic state made of an isothermal and stratified resting atmosphere where the temperature $T_0$ is a constant and where the potential temperature $\theta_0(p) = T_0 \: (p/p_{00})^{\kappa}$ varies with $p$.
It ensues from Eq.(\ref{eq_8})  that the pseudo-energy can be written as ${\cal A} = E_k + A_T$, where
\begin{equation}
\boxed{\;\;
A_T
\; = \; 
\iiint_{\cal M} 
     c_p \; 
    \left[ \;   
      ( T - T_0 ) - T_0 \: \ln(T/T_0)
    \; \right] 
\; dm
\; = \; 
\iiint_{\cal M} 
     c_p \; T_0 \;\:
    {\cal F}(\,T/T_0 - 1 ) 
\; dm
\: .
\;\;}
\label{eq_10}
\end{equation}
It is the temperature component of the global available enthalpy $A_h = A_T + A_p$ given by Eq.(\ref{eq_9}), where
\begin{equation}
\boxed{\; \;
{\cal F}(X) \; = \; X - \ln(1 + X)
\;\;}
\;\;\;\;
\mbox{and}
\; \;
\boxed{\;\;
 X \; = \; \frac{T}{T_0} \: - \: 1
\;\;}
\: .
\nonumber
\end{equation}
Equation (\ref{eq_10}) is the primary expression obtained by Thomson (1853) in thermodynamics when he studied the maximum work obtainable from an unequally heated space.\footnote{\color{blue} 
Again, see Marquet (1991) \url{http://arxiv.org/abs/1402.4610} {\tt arXiv:1402.4610 [ao-ph]}.}
This equation can be approximated using ${\cal F}(X) \approx X^2/2$ for small $|X|$,  in order to obtain the two meteorological formulations of Pearce (1978) and Blackburn (1983): $a \approx c_p \: (T - T_0)^2 /(2 \: T_0)$. 
Therefore the pseudo-energy of Shepherd also leads to these two meteorological results.

Another isothermal reference state $T_0=Cste$ have been used by Andrews (1981) to define the {\it potential energy\/} \footnote{\color{blue} The potential energy of Andrews (1981) was not mentioned in Marquet (1995). This paragraph is puiblished in Marquet (2003b).} for a perfect gas by $\Pi = {\Pi}_1(p/p_0)+ {\Pi}_2(\theta/{\theta}_0)$, where the notation $\Pi$ retained in Andrews (1881) does not represent the Exner's function.
The two parts ${\Pi}_1$ and ${\Pi}_2$ are local and positive definite everywhere.
The second part can be written as  ${\Pi}_2(\eta)= c_p \: T_0 \: {\cal G}(\eta)$, with ${\cal G}(\eta) = \exp(\eta) -1 - \eta$ and $\eta=\ln(\theta/{\theta}_0)$.
It is thus equal to ${\Pi}_2(X)= c_p \: T_0 {\cal F}(X)$, where $X=\theta/{\theta}_0-1$ and the function ${\cal F}(X) = X - \ln(1+X)$ is the same function used to define $a_T=c_p T_r {\cal F}(T/T_r-1)$.
The temperature component of the {\it potential energy\/} of Andrews (1981) is thus obtained as a special case of the pseudo-energy of Shepherd.

The formulation of McHall (1990) is not so easy to derive from Eqs.(\ref{eq_7})  or (\ref{eq_8}). 
One of the reasons could be the fact that he makes use of two conservation laws whereas other studies involve only one conservation law. 
McHall takes the total entropy and the integral of the potential temperature constant and he searches for states of minimum possible enthalpy. 
Lorenz and Dutton, for instance, are rather concerned with conservation of mass (between two isentropes or the global mass, respectively), and they search for states of minimum enthalpy (Lorenz) and maximum entropy (Dutton).

It is not easy either to discover the connection between the pseudo-energy of Shepherd and another form of static entropic energy defined by Dutton, which is the specific exergy of Karlsson (1990) and Gibbs (1879).
The basic states of Dutton and Karlsson are the same stratified and isothermal atmosphere that have been considered above to find the results of Thomson, Pearce and Blackburn.
The main difference is that the theories of Dutton and Karlsson can deal with a hydrostatic or non-hydrostatic atmosphere associated with a hydrostatic reference basic state where the temperature is constant at $T_0$. 
Karlsson defines the specific availability function by $c_v \: T_0 \: {\cal F}(T/T_0 - 1) + R \: T_0 \: {\cal F}( \rho_0(z)/\rho - 1)$ which is positive and,  from ${\cal F}(X) \approx X^2/2$, approximately doubly quadratic with respect to $T/T_0 - 1$ and $\rho_0(z)/\rho - 1$. 
It is thus the integrand of one of the global integrals introduced by Dutton, but it is not equal to the available enthalpy (\ref{eq_9}) which is the hydrostatic pseudo-energy for the isothermal reference state.
Therefore, it would still be interesting to derive the pseudo-energy for a non-hydrostatic atmosphere and to make the comparison with the results of Dutton and Karlsson.

One of the remaining problems is that, at first sight, the available enthalpy (\ref{eq_9}) is not positive definite. 
Of course, from ${\cal F}(X) \approx X^2/2$ the first component $A_T$ is positive and of quadratic order with respect to the disturbance amplitude $T - T_0$, and this is necessary by virtue of the general method of Shepherd.
But the integrand of the pressure component $A_p$ is not of any definite sign. 
This problem is solved in the available enthalpy approach of Marquet (1991) by choosing the reference pressure $p_r$ so that the global integral $A_p$ cancels out.
The definition $p_r = p_{00} / e$ and $A_p=0$ are thus coherent with the positive and quadratic global function $A_h = A_T$.\footnote{\label{footnote_ap_Margules}{\color{blue} 
A criticism often reported is that $a_p(p)$ is not a positive quantity.
However, it is worth noting that $R \: T_r \: \ln(p/p_r)$}
{\color{blue} can be rewritten in terms of the derivative of a function ${\cal H}$ defined by ${\cal H}(X)=(1+X)\:\ln(1+X) - X$ and $X = p/p_r -1$, leading to $a_p = R \: T_r \: p_r \: d/dp[{\cal H}(X)]$.
This function ${\cal H}(X)$ is defined for $p>0$, i.e. for $X>-1$ as for ${\cal F}(X)$.
The leading approximation of ${\cal H}(X)$ for small $|X|$ is $X^2/2$ as for ${\cal F}(X)$.
It is easy to demonstrate that ${\cal H}(X)$ is positive, equal to zero only for $p=p_r$ and equal to $1$ for $p=0$. 
Accordingly, the global integral of $a_p$ (i.e. $A_p$) is roughly proportional to ${\cal H}(p_s/p_r-1)\approx (p_s-p_r)^2/(2\: p_r^2)$.
Moreover, accurate computations show that $A_p$ depends on the horizontal variance of surface pressure $\overline{(p_s-\overline{p_s})^2/ (\overline{p_s})^2}$.
This result was already derived in Margules (1901), in the paper dealing with ``The mechanical equivalent of any given distribution of atmospheric pressure, and the maintenance of a given difference in pressure''. 
This paper was published some years before the more famous one about the ``energy of storms'' (1903-05).}
}

The two hydrostatic formulations (\ref{eq_9}) and (\ref{eq_10}) are thus the same on a global stage providing that $p_r = p_{00} / e$, but they are associated with two different local available energies: $a_T + a_p$ for  (\ref{eq_9}), only $a_T$ for (\ref{eq_10}). 
The specific version of $A_T$ and $A_p$ are respectively $a_T = c_p \: [ \: (T - T_r) - T_r \: \ln(T/T_r) \: ]$ and $a_p = R \: T_r \: \ln(p/p_r)$. 

In fact the term $a_p$ is important in local budgets in order to balance the conversion term $- R \: \omega\: T_r/p$ between $a_T$ and $a_p$ ($\omega$  is the vertical velocity). 
With Eq.(\ref{eq_10}), the local budget of $a_T$ cannot be closed with the kinetic-energy equation and $- R \: \omega\: T_r/p$  remains as an extra conversion term. 
This problem does not arise with the specific available enthalpy function $a_h = a_T + a_p$ which is at the same time consistent on local and global points of view.

% -----------------------
 \section{CONCLUSION.} % (Section 4)
% -----------------------
\label{section_4}

This note establishes a close connection between the generalized available potential (pseudo)-energy of Shepherd (1993) and almost all other forms of availability functions used in meteorology (Shepherd had already derived Lorenz's approach of APE (1955) as a small-amplitude limit). 

First, an alternative formulation is given by Eq.(\ref{eq_8}) which appears to be suitable in order to deal with all various isothermal basic states.

Then, it is shown that the global available enthalpy (Marquet, 1991) and the global hydrostatic results of Dutton (1973)  and  Pichler  (1977)  correspond to a basic state made of an isothermal and isobaric parcel at $T_r$ and $p_r$. 
As for the available energies of Thomson (1853), Pearce (1978) and Blackburn (1983), they correspond to the basic state made up of an isothermal and stratified resting atmosphere at $T_0(p)$ with a variable pressure.

The global and local available enthalpy formulations ($A_h$ and $a_h$) are different from Lorenz's APE results, which are more familiar for meteorological purposes.
$A_h$ and $a_h$ are  also different from the theories of Dutton and Pearce. 
It was, therefore, important to demonstrate these connections via the general concept of pseudo-energy of Shepherd, and to prove that the two special cases of basic states described above generate the two families of concepts used in meteorology: the available potential energy of Lorenz on the one hand, and the various exergy functions of Dutton, Pearce and Marquet on the other hand.

The specific exergy functions such as $a_h$ or $T_0\:\sigma$ have already been studied and applied on a local point of view. 
The meteorological properties of the specific static entropic energy $T_0\:\sigma$ were first investigated by Pichler (1977). 
As for the properties of the specific available enthalpy $a_h$, they have been explored by Marquet (1991, 1993, 1994, 2003a,b).\footnote{\color{blue} The papers published in 2003a,b deal with the definition and the use of an available enthalpy cycle. 
The paper published in 1993 deals with the definition of a {\it moist-air\/} enthalpy cycle.}

It is proven by Marquet (1994) that a hydrostatic available enthalpy cycle can be rigorously derived for an isobaric layer of a limited-area domain. 
It appears that a Lorenz-like cycle is embedded in a more general one where various boundary fluxes are associated with each energy reservoir. 
There are also three large terms all depending on the isobaric average of the vertical velocity ($\overline{\omega}$) which is equal to zero only for the global atmosphere. 
In fact there are two orders of magnitude between these large terms and the other one of the cycle, but it is important to notice that the Lorenz-like cycle is somewhat protected from these large terms because they only touch
the static stability component $a_S$ (depending on the vertical variations of $T - T_r$), and the kinetic energy of the mean wind $[\: {(\overline{u})}^2 + {(\overline{v})}^2 \:]/2$. It turns out also that the boundary fluxes are important when considering the energetic of a baroclinic wave on a limited area.

All these results show that it was indeed important to arrive at a local definition of the concept of availability since new properties can be demonstrated. The local pseudo-energy of Shepherd seems to generalize all previous thermodynamical and meteorological exergy-like functions.

\vspace{5mm}
\noindent{\Large\bf Acknowledgements}
\vspace{2mm}

Thanks are due to J. P. Lafore for the personal communication of Shepherd's work.

%-------------------------------------------------------------------------
%    REFERENCES
%-------------------------------------------------------------------------

%\begin{thebibliography}
%\newpage
\vspace{5mm}
\noindent{\Large\bf References}
\vspace{2mm}

\noindent{$\bullet$ Andrews,~D.} {1981}.
{A note on potential energy in a stratified compressible fluid.
{\it J. Fluid Mech.\/}
{\bf 107,}
p.227--236.}

\noindent{$\bullet$ Blackburn,~M.} {1983}.
{\it An energetic analysis of the general atmospheric circulation\/}.
Thesis of the department of Meteorology. University of Reading, UK.

\noindent{$\bullet$ Dutton,~J.~A.} {1973}.
{The global thermodynamics of atmospheric motion.
{\it Tellus.\/}
{\bf 25,} (2),
p.89--110.}

 \noindent{$\bullet$ Gibbs,~J.~W.} {1873}.
{A method of geometrical representation
of the thermodynamic properties of substance
by means of surfaces.
{\it Trans. Connecticut Acad.\/}
{\bf II}: p.382--404.
(Pp 33--54 in Vol. 1 of
{\it The collected works of J. W. Gibbs,\/}
1928.
Longmans Green and Co.)} 

\noindent{$\bullet$ Karlsson,~S.} {1990}.
{{\it Energy, Entropy and Exergy in the atmosphere.\/}
Thesis of the Institute of Physical Resource Theory.
Chalmers University of Technology.
 G\"oteborg, Sweden.}

\noindent{$\bullet$ Lorenz,~E.~N.} {1955}.
{Available potential energy and the 
 maintenance of the general circulation.
{\it Tellus.\/}
{\bf 7,} (2),
p.157--167.}

\noindent{$\bullet$ Margules,~M.} {1901}.
{The mechanical equivalent of any given distribution of 
atmospheric pressure, and the maintenance of a given 
difference in pressure. 
{\it Smithsonian Miscellaneous collections.}
{\bf 51,} (4): 501--532, 1910
{\it (Translation by C. Abbe of a lecture read at the 
meeting of the Imperial Academy of Science, Vienna, 
July, 11, 1901, commemorating the Jubilee of the 
Central Institute for Meteorology and Terrestrial 
Magnetism)}.}

\noindent{$\bullet$ McHall, ~Y.~L.}, {1990}.
{Available potential energy in the atmospheres.
{\it Meteorol. Atmos. Phys.\/}, 
{\bf 42}, 
39--55.}

\noindent{$\bullet$ McHall, ~Y.~L.}, {1991}.
{Available equivalent potential energy in moist atmospheres.
{\it Meteorol. Atmos. Phys.\/}, 
{\bf 45}, 
113--123.}

\noindent{$\bullet$ McIntyre, ~M.~E. and Shepherd,~T.~G.}, {1987}.
{An exact local conservation theorem for finite-amplitude 
 disturbances to non-parallel shear flows, with remarks 
 on Hamiltonian structure and Arnol'd's stability theorems.
{\it J. Fluid Mech.\/}, 
{\bf 181}, 
527--565.}

\noindent{$\bullet$ Marquet~P.} {1991}.
{On the concept of exergy and available
enthalpy: application to atmospheric energetics.
{\it Q. J. R. Meteorol. Soc.}
{\bf 117}:
449--475.
\url{http://arxiv.org/abs/1402.4610}.
{\tt arXiv:1402.4610 [ao-ph]}}

\noindent{$\bullet$ Marquet~P.} {1993}.
{Exergy in meteorology: definition and properties
of moist available enthalpy.
{\it Q. J. R. Meteorol. Soc.}
{\bf 119} (511) :
567--590.} 

\noindent{$\bullet$ Marquet,~S.} {1994}.
{{\it Applications du concept d'exergie \`a  l'\'energ\'etique de l'atmosph\`ere.
Les notions d'enthalpies utilisables s\`eche et humide.\/}
 Thesis of the University of Paul Sabatier, Toulouse, France.

\noindent{$\bullet$ Marquet,~P.} {1995}.
{On the concept of pseudo-energy of T. G. Shepherd.
{\it Q. J. R. Meteorol. Soc.}
{\bf 121}:
455--459.}

\noindent{$\bullet$ Marquet,~P.} {2003a}.
{The available-enthalpy cycle. I: 
Introduction and basic equations.
{\it Q. J. R. Meteorol. Soc.}
{\bf 129}:
2445--2466.}

\noindent{$\bullet$ Marquet,~P.} {2003b}.
{The available-enthalpy cycle. II: 
Applications to idealized baroclinic waves.
{\it Q. J. R. Meteorol. Soc.}
{\bf 129}:
2467--2494.}

\noindent{$\bullet$ Pearce,~R.~P.} {1978}.
{On the concept of available potential energy.
{\it Q. J. R. Meteorol. Soc.\/}
{\bf 104},
p.737--755.}

\noindent{$\bullet$ Pichler,~H.} {1977}.
{Die bilanzgleichung f\"ur die statischer 
entropische Energie der Atmos\-ph\"are.
{\it Arch. Met. Geoph. Biokl.\/}, 
Ser.A, 
{\bf 26},
p.341--347.}

\noindent{$\bullet$ Rant,~Z.} {1956}.
{Exergie, ein neues Wort f\"{u}r ``Technische 
 Arbeitsf\"{a}higkeit''.
{\it Forsch. Ing. Wes.\/}, 
{\bf 22},
p.36--37.}

\noindent{$\bullet$ Shepherd,~T.~G.}, {1990a}.
{A general method for finding extremal states of 
Hamiltonian dynamical systems, with applications 
to perfect fluids.
{\it J. Fluid Mech.\/}, 
{\bf 213}, 
573--587.}

\noindent{$\bullet$ Shepherd,~T.~G.}, {1990b}.
{Symmetries, conservation laws, and Hamiltonian 
structure in geophysical fluid dynamics.
{\it Adv. Geophys.\/}, 
{\bf 32}, 
287--338.}

\noindent{$\bullet$ Shepherd,~T.~G.} {1993}.
{A unified theory of available potential energy.
{\it Atmosphere Ocean.\/}
{\bf 31}, (1),
p.1--26.}

\noindent{$\bullet$ Shepherd,~T.~G.} {2003}.
{Hamiltonian dynamics. In Encyclopedia of Atmospheric Sciences (J.R. Holton et al., eds.), 
{\it Academic Press.}, pp. 929--938}

\noindent{$\bullet$ Thomson,~W.} {1853}.
{On the restoration of mechanical energy 
from an unequally heated space.
{\it Phil. Mag.\/}
{\bf 5,} 30, 4e series,
p.102--105.}

\noindent{$\bullet$ Thomson,~W.} {1879}.
{On thermodynamic motivity. 
{\it Phil. Mag.\/}
{\bf 7,} 44, 5e series,
p.346--352.} 

\end{document}